# Deep learning-based fully automatic segmentation of wrist cartilage in MR images

Ekaterina Brui[1], Aleksandr Y. Efimtcev[1,2], Vladimir A. Fokin[1,2], Remi Fernandez[3], Anatoliy G. Levchuk[2], Augustin C. Ogier[4], Alexey A. Samsonov[5], Jean. P. Mattei[4, 6], Irina V. Melchakova[1], David Bendahan[4,] Anna Andreychenko[1,7].

[1] University of Information Technology Mechanics and Optics, International Research Center Nanophotonics and Metamaterials, 199034 S.-Petersburg, Russia
[2] Federal Almazov North-West Medical Research Center, 197341 S.-Petersburg, Russia
[3] APHM, Service de Radiologie, Hôpital de la Conception, Marseille, France
[4] Aix-Marseille Universite, CNRS, Centre de Résonance Magnétique Biologique et Médicale, UMR 7339, Marseille, France
[5] University of Wisconsin-Madison, Department of Radiology, Madison, WI 53705-2275 USA
[6] Assistance Publique Hôpitaux de Marseille, Institut de l'appareil locomoteur, Service de Rhumatologie, Hôpital Sainte Marguerite, Marseille, France
[7] Research and Practical Clinical Center of Diagnostics and Telemedicine Technologies, Department of Health Care of Moscow, Moscow, Russia

**Corresponding Author's contact information**
Ekaterina Brui, e-mail: e.brui@metalab.ifmo.ru, katya.bruy@gmail.com, tel: +7 9618112918, address: 199034, Russia, Saint Petersburg, Birjevaja line V.O., 14

**Keywords**
Cartilage < Musculoskeletal < Applications, Quantitation < Postacquisition Processing < Methods and Engineering, Human study < Musculoskeletal < Applications

**Abstract**
The study objective was to investigate the performance of a dedicated convolutional neural network (CNN) optimized for wrist cartilage segmentation from 2D MR images. CNN utilized a planar architecture and patch-based (PB) training approach that ensured optimal performance in the presence of a limited amount of training data. The CNN was trained and validated in twenty multi-slice MRI datasets acquired with two different coils in eleven subjects (healthy volunteers and patients). The validation included a comparison with the alternative state-of-the-art CNN methods for the segmentation of joints from MR images and the ground-truth manual segmentation. When trained on the limited training data, the CNN outperformed significantly image-based and patch-based U-Net networks. Our PB-CNN also demonstrated a good agreement with manual segmentation (Sørensen–Dice similarity coefficient (DSC) = 0.81) in the representative (central coronal) slices with large amount of cartilage tissue. Reduced performance of the network for slices with a very limited amount of cartilage tissue suggests the need for fully 3D convolutional networks to provide uniform performance across the joint. The study also assessed inter- and intra-observer variability of the manual wrist cartilage segmentation (DSC=0.78-0.88 and 0.9, respectively). The proposed deep-learning-based segmentation of the wrist cartilage from MRI could facilitate research of novel imaging markers of wrist osteoarthritis to characterize its progression and response to therapy.

**Abbreviations used**
CNN - convolutional neural network; DSC - Sørensen–Dice similarity coefficient; OA –osteoarthritis; CSA - cross-sectional area, VIBE - Volumetric Interpolated Breath-hold Examination, O1 – observer 1, O2 – observer 2, O3– observer 3, O4 – observer 4.

**Funding information**
This work was supported by the Russian Science Foundation (Grant No. 18-79-10167). This project has received funding from the European Union's Horizon 2020 research and innovation programme under grant agreement No. 736937. This work was partially supported by NIH (R01EB027087).
**The authors declares that there is no conflict of interest**
**Word count: 3868**

# INTRODUCTION

MRI is a versatile tool for the detection of morphological and compositional cartilage abnormalities in degenerative diseases of joints[1]. MRI-based measurements of a joint space narrowing[2,3,4] have been utilized to assess cartilage degradation in multiple locations including knee[5] and wrist[6]. MRI has also been applied to quantify other morphometric features including cartilage cross-sectional area (CSA)[7] and cartilage volume[8,9]. More recently, several quantitative MRI approaches have been proposed to assay proteoglycan and collagen components of the cartilage matrix[10,11,12,13]. It has been suggested that such biomarkers could be used for disease detection and treatment monitoring.

Analysis of structural and quantitative MRI data requires an accurate cartilage segmentation, whose automation for routine applications is challenged by the presence of other tissues with similar MR contrast (e.g, muscles, skin, edematous tissues). On that basis, a manual segmentation is considered as the gold-standard in cartilage assessment applications[8,14,15,16,17]. However, the manual segmentation is a highly time-consuming and tedious task and its reliability can be hampered by the inter-operator variability. To improve the speed and consistency of the cartilage segmentation, a wide variety of computer-assisted approaches has been proposed including semi-automated[18,19] and fully automatic[20,21,22,23] segmentation methods. These approaches provide a very fast and reliable segmentation with a moderate penalty on the segmentation accuracy i.e. with Sørensen–Dice similarity coefficient (DSC)[24] approaching 0.80[20]. Most recently, convolutional neural networks (CNN)[25,26,27,28] have been successfully applied for the segmentation of knee MR images. The methods demonstrated improved DSC values of 0.82[25] and 0.88[26] for planar and U-Net[29] architectures, respectively. This highlighted an initial promise of the machine learning approaches for the fully automated segmentation of complex anatomical structures.

MRI has proven to be a promising approach for wrist OA evaluation demonstrating higher sensitivity to moderate changes of OA as compared to X-ray based assessment[30]. MRI-based wrist joint assessment is potentially more suitable to follow changes over time and/or to assess the efficiency of therapy than CT arthrography given that harmful ionizing radiation and injection of a contrast material into the joint space are a part of the CT imaging procedure[31]. However, while most of automated segmentation methods have been developed for knee MRI, only a few[7,9] techniques have been optimized for the wrist MR image segmentation, likely due to the more complex anatomy of the wrist joint. The automatic segmentation of the wrist joint cartilage from MRI images could facilitate research of novel imaging biomarkers of wrist OA and to characterize its progression and response to therapy.

Given the outstanding performance of CNN-based approaches for the segmentation of the knee structures[25,26], we hypothesized that deep-layered CNNs could be valuable in the design of an automatic segmentation of wrist cartilage. Therefore, in the present work, we developed, optimized, and evaluated a CNN-based method for a fully automatic segmentation of wrist joint cartilage. The network was trained using the manual labels produced by experienced radiologist and compared with several representative CNN-based methods[26,32.]

# PATIENTS AND METHODS

**Subjects**

The study was approved by the local ethics committee. Eleven subjects were enrolled into the study after obtaining the written informed consent. These included eight healthy volunteers (no previous wrist trauma, six males and two females, age range 23-38, mean 29.6) and three patients (two 63 and 77-year-old females with confirmed OA diagnosis, and one 62-year-old female with articular pain). All data were acquired in the dominant wrist.

**MR-imaging**

MR images were acquired at 1.5T Magnetom Espree system (Siemens GmbH, Erlangen, Germany). The same wrist was scanned twice, first with a conventional "birdcage"-type transmit/receive extremity coil and then with a home-made wireless coil providing a higher signal-to-noise ratio (SNR)[33]. In two subjects, one of the scans was not completed due to technical or cooperation reasons thereby bringing the total number of MRI scans to 20. 3D coronal T1-weighted gradient echo (VIBE - Volumetric Interpolated Breath-hold Examination) images with water-selective excitation for fat suppression were acquired to achieve an optimized contrast-to-noise ratio for the cartilage[7]. The relevant parameters were: TR/TE = 18.6/7.3 ms, flip angle = 10˚, FOV = 97x120 mm$^2$, matrix size = 260x320, voxel size = 0.37x0.37x0.5 mm$^3$, number of coronal slices = 88. Total acquisition time was 6 min.

**Data preparation**

3D regions of interest (ROIs) were manually outlined to encompass all cartilage tissues in the acquired volumes. Every other slice was selected from all 3D ROIs to form an intermediate dataset containing 420 images with cartilage. Then, the dataset was augmented by an additional set of 140 slices that did not contain cartilage tissue, with a final dataset comprising 560 images.

*Dataset labeling*

Image processing was performed using MATLAB (MathWorks, Natick, Massachusetts). Cartilage tissue was segmented by an expert radiologist (O1 - V.F.) using a software-assisted manual approach. In detail, the wrist joint ROI delineated for each slice by manual contouring (Fig. 1 a) was first roughly segmented by intensity thresholding, with threshold value optimized by the observer in iterative fashion on per slice basis. Next, the resulting binary masks (Fig. 1 b) were manually corrected to ensure that only cartilage pixels were included into the labels.

*Data splitting and CNN training approaches*

The labeled images were split into several subsets to be used for training (n=260), development (n=20), testing (n=260), and method validation (n=20) stages (Fig. 2). We utilized an "hold-out" training approach, in which training and development datasets included data from healthy volunteers #1 to #5 (MRI scans #1 to #10) and test dataset included data from subjects #6 to #11 (MRI scans #11 to #20). For each subject, the medial coronal slice chosen from the 3D dataset was used for the validation of our method and comparison to the manual one. To ensure these images are unseen by CNN, these slices were excluded from the training, development, and test datasets.

To assess the CNN robustness with respect to the anatomical heterogeneity of the training dataset, several CNN variants were trained using different training subsets containing: 1) 5%, 10%, 20%, 33%, 50%, 66% of the total number of slices (n=260) randomly selected from the training dataset; 2) images of particular subjects, starting from subject #1 and then subsequently adding other subjects' data.

Additionally, we performed cross-validation studies for three patients and three healthy volunteers included in the testing dataset to estimate the best achievable network performance in the presence of a limited number of subjects. In these studies, the network was trained using 6 different samples of 10 subjects (6-fold analysis) and tested on the remaining subjects (from #6 to #11).

For each CNN instance, the performance was evaluated by comparing the CNN-based and the manual segmentations using DSC values for 3D images or for planar slices independently. The signal-to-noise ratio was measured as the ratio between the cartilage tissue signal and the standard deviation of the noise within a signal-free area.

**CNN-assisted segmentation**

*Patch-based (PB) CNN architecture*

The design of our network was based on our preliminary experiments, which indicated that the state-of-the-art U-Net-based CNN[26] did not perform satisfactory, likely as a result of the small size of our training dataset (260 images) and the large number of trainable parameters ($2,8*10^6$). On that basis and in order to minimize the risk of overfitting, we selected a planar network architecture with a smaller number of trainable parameters and a patch-based (PB) training approach, which proved to be adequate for limited amount of annotated data[32,34,35]. The network parameters including the number of convolutional layers, the number and the size of filters were optimized by the grid search. The final PB-CNN architecture (Figs. 3 and A1) had five convolutional layers with 44 filters of size 3x3. The patch size was further optimized for this architecture. Gaussian noise and drop-out regularization layers were added to reduce the generalization error and to minimize risk of overfitting. The network was trained during 20 epochs with a 20000 batch-size. Ten percent of training data were used for the training validation (i.e., for the calculation of loss value). The stopping criterion during the training phase was defined as the absence of loss value decrease for 5 epochs. A RMSProp optimizer (https://keras.io/optimizers/) with a default learning rate value of 0.001 was used.

*PB-CNN input and output*

We used a sliding-window approach[25] to select patches surrounding the pixels to be classified. In detail, for each pixel of interest, a 28x28 patch centered on the pixel was applied to provide a network input. For each image, the network output a probability map, which was thresholded to obtain cartilage binary masks. The threshold value was optimized during the development stage to maximize DSC coefficient with respect to the development dataset. Overall, datasets for method development and validation contained $1.3*10^6$ patches. For the "hold-out" approach, both training and testing datasets contained a total of $17.6*10^6$ patches.

For the subject-based cross-validation, the training and testing samples were varying for every step of the 6-fold analysis. The number of patches in the training dataset varied from $15.9*10^6$ to $16.7*10^6$, depending on how many times (1 or 2) each subject had been scanned. The networks were tested on one or two separate 3D image ($1.8*10^5$ patches) of subjects that were not included in the trained dataset.

*State-of-the art neural networks*

We compared our proposed network with several alternative architectures such as image-based[26] and patch-based[32] U-Net CNNs trained and tested with the "hold-out" approach. These CNNs are detailed in Appendix 1 (Fig. A2 and Fig. A3).

*Hardware and software*

The training was performed on a server with four processors (Intel Xeon E5-4617 2.90 GHz) and 512 Gb RAM. To provide realistic estimates of the network execution speed, all methods were tested on a PC with more common characteristics (Intel Core i5-7640X processor, 32 Gb of RAM). CNNs were built using Python 3.6.4, TensorFlow 1.7.0, and Keras 2.1.5 open-source neural network library.

**Data analysis**

*Reproducibility of the manual segmentation procedure*

A reproducibility of the manual segmentation procedure used to create labels from the training dataset was assessed from the segmentation results for the method validation dataset obtained by three observers, each with more than 10 years of experience in musculoskeletal segmentation. The first observer (O1 – V.F.) segmented the wrist cartilage twice. The segmentation sessions were separated by one week. The segmentation results from the first session were considered as the ground truth for the purposes of the method evaluation. Two other experts (O2 - A.E., O3 - R.F.) segmented images once.

Cartilage CSA was calculated as a product of the number of pixels within the binary mask and the pixel area (0.37x0.37 $mm^2$). For each observer, the averaged cartilage CSA was calculated among the method validation dataset and compared statistically using Student's t-tests. To determine the variability of the manual segmentation procedure, the inter- and intra-observer $DSC=2|X \cap Y|/(|X|+|Y|)$ [24] values were calculated, where X is a binary mask segmented by O1 in the first session, and Y – segmented either by O1 in the second session (intra-observer study) or by O2 and O3 (inter-observer study).

*PB-CNN-based cartilage segmentations*

Performances of the trained networks were evaluated based on a comparisons with the ground truth i.e. the manual segmentation as described above. For the "hold-out" approach, DSC value was calculated independently for each 3D volume from the test dataset and then averaged. The layer-to-layer analysis of the segmentation accuracy for the developed PB-CNN was performed according to two stages. Four cartilage zones along the slice selection direction in each 3D image were initially identified. Then the cartilage volume in each zone was normalized with respect to the volume of the medial slice in the corresponding 3D image i.e. the image containing the largest amount of cartilage, which was assumed as 100%. The identification of the cartilage zones was designed to account for differences in wrist joint thickness among different volunteers. Zone #1 encompassed slices with no cartilage, zone#2 - slices with a relative amount of cartilage up to 33%, zone#3 - slices with an intermediate amount of cartilage - from 34 to 66%, and zone#4 - slices with an amount of cartilage ranging from 67 to 100%. An averaged DSC index was calculated for each zone of the 3D images.

In the cross-validation analysis, the trained networks were tested on 3D images (1 or 2) of the subjects not included in the training dataset. Corresponding DSC values were calculated for each 3D image as a whole and on the layer-to-layer basis.

The evaluation of the network trained using the "hold-out" approach was performed as follows: DSC was calculated individually for 20 medial coronal slices from the method validation dataset and compared with the human inter- and intra-observer study results. A radiologist with more than 10 years of experience in musculoskeletal MRI (O4 – A.L.) performed a 10-point visual evaluation of the PB-CNN-based segmentations. Several factors of a segmentation quality were assessed; their presence in the cartilage mask led to a reduction of the initial score of 10 by:

*3 points*
- segmentation of bone tissue pixels
- segmentation of pixels of pathological zones in bones
- segmentation of pixels out of the wrist joint

*2 points*
- significant amount of non-segmented cartilage tissue on articular surfaces of wrist joint

*1 point*
- non-segmented cartilage tissue in three or more bones articulation
- deviations of the thickness of segmented cartilage
- segmented non-cartilage pixels in wrist joint.

## RESULTS
**Manual segmentation procedure**

The averaged cartilage CSA determined by the first observer and considered as the ground truth was $237.6 \pm 39.8$ mm$^3$. The corresponding result obtained by O2 ($240.1 \pm 38.1$ mm$^3$) was not significantly different (p>0.05), whereas O3 reported a significantly higher CSA ($269.3 \pm 39.1$ mm$^3$, p=0.015).

The average rate for the manual segmentation was 20 slices/hour (i.e. 5 min per slice). Intra-observer DSC was 0.90±0.04 whereas inter-observer DSC values were lower (0.88±0.04 for O1 vs O2 and 0.78 ± 0.06 for O1 vs O3). The corresponding statistics for the manually performed segmentation procedure are summarized in Table 1. While the SNR of images differed from one subject to another due to the differences related to coil load, the SNR was systematically higher (p>0.05) with the home-built coil. The paired t-test (p > 0.05) for the data of subjects scanned twice did not provide any significant difference in the segmentation accuracy both for intra- and inter-observer studies.

**CNN-assisted segmentation**

*Performances of CNNs and Sensitivity analysis*

Figure 4 shows the results of the sensitivity analysis from the "hold-out" training/testing stage. DSC value for the development dataset raised continuously up to DSC = 0.86 when the data of each subject was subsequently added to the training dataset. When the slices for training were randomly selected from the whole dataset in different proportions, DSC raised in a similar manner. The DSC improvement was less than 6% while the training sample size increased from half to full training dataset (i.e., twice). Training duration on a full training dataset was 74.4 hours. The segmentation time was 15 s per slice.

Figure 4 additionally demonstrates the performance of state-of-the-art networks on a development dataset in comparison with the proposed here PB-CNN. For the classical image-based U-Net CNN, DSC was much lower (0.64). The segmentation time was 0.75 s per slice. For the patch-based variant of the U-Net CNN, the training time per epoch was 23 times longer than for our PB-CNN, which deemed it infeasible to train the network in an acceptable time. Therefore, the network was trained using a reduced training patch database (~1/20 of the full one) to stay within the feasibility limit. The corresponding DSC value was 0.44 and the segmentation time was 2.05 min per slice.

*PB-CNN performance validation*

Results to the CNN-based cartilage segmentation procedure accuracy for the method validation phase are summarized in Table 1. The averaged DSC was 0.81±0.05. The averaged cartilage CSA (266.5± 34.3 mm$^3$) was significantly higher (p = 0.019) than the corresponding value quantified by the expert (237.6 ± 39.8 mm$^3$). According to the paired t-test (p > 0.05), the segmentation accuracy was not influenced by the coil selection.

Examples of CNN-based segmentations of different accuracies are displayed in Figure 5. The CNN-based segmentation of the validation dataset was qualitatively assessed by an independent radiologist who scored 7.00±1.51 for controls and 4.60±1.14 for patients (Table 1). The scores were consistent with the average DSC values for these cartilage masks (0.82±0.03 and 0.76±0.04, for healthy volunteers and patients, respectively).
.
*Performance of PB-CNN across 3D volume*

Table 2 shows DSC scores for a layer-to-layer experiments, which involved analysis of different plane locations in 3D. DSC values averaged across the 3D images of the testing dataset were 0.69 ± 0.06 for the whole group, 0.73 ± 0.03 for the healthy subjects, and 0.65 ± 0.05 for the patients. The averaged DSC value in the cross-validation analysis was 0.70±0.05 (0.73±0.02 and 0.67±0.05 for controls and patients, respectively). The layer-based analysis showed that the DSC was lower for the slices located far away from the medial cross-section both for "hold-out" and cross-validation (in brackets) tests: zone#1: 0.21±0.21 (0.25±0.20), zone#2: 0.60±0.09 (0.61±0.09), zone#3: 0.63±0.06 (0.65±0.05), zone#4: 0.73±0.05 (0.74±0.05).

## DISCUSSION

The aim of the present study was to investigate the performance of a dedicated CNN for the segmentation of the wrist cartilage from structural MR images, to compare the corresponding results with manual and existing CNN-based segmentation approaches, and to assess the dependence of the network's performance on the amount and heterogeneity of the training data.

Our results demonstrated that the presented PB-CNN architecture significantly outperformed the classical image-based U-Net in the wrist cartilage segmentation task (DSC=0.86 and 0.64, respectively). This improvement came at the expense of the computational time (15 s for our PB-CNN vs 0.75 s for the image-based U-Net CNN). The decreased accuracy of the image-based network may be explained by the relatively low number of training samples, which, however, did not result in reduced performance of our patch-based method. Our PB-CNN also outperformed significantly the patch-based U-Net architecture under identical training conditions (Fig. 4). Interestingly, our network was much faster than PB-U-Net, which may be explained by much lower number of parameters in our planar PB-CNN and the need for full patch mask for PB-U-Net training. It should be noted that we utilized basic U-net architectures that were not fine-tuned for the wrist joint cartilage segmentation on MR images. Overall, our PB-CNN provided a fast and reliable segmentation of wrist cartilage in MR image.

The performance of our PB-CNN agreed well with that of the manual segmentation, as evidenced by the comparable DSC values (0.81±0.05) with those from the inter-observer assessement (0.88±0.04 for O1 vs O2 and 0.78 ± 0.06 for O1 vs O3 in medial slices). At the same time, it was lower as compared to the intra-observer study (0.90±0.04) thereby indicating that the PB-CNN did not fully reproduce the manual segmentation strategy of Observer 1 (one must keep in mind that the same observer (O1) segmented the wrist cartilage for training and for the intra-observer study). Yet, the DSC values were comparable to those previously reported for CNN-assisted knee cartilage segmentation (0.82-0.88)[25,26], which demonstrates a promise of CNN-based methods for more challenging wrist cartilage segmentation.

Our network demonstrated robustness regarding many anatomical structures and joint abnormalities that have appearances or contrast similar to cartilage, as illustrated by representative examples in Figure 5 (e.g., a vessel in the capitate bone of a healthy volunteer (Fig. 5 a) and cyst and cortical bone erosions in patients Figs. 5 e, f). At the same time, the analysis indicated an elevated number of false positive pixels in images of patients, mostly due to misclassification of skin tissue as cartilage (Fig. 5 d). Our cross-validation study demonstrated that PB-CNN segmentation accuracy of patients' data benefited from larger heterogeneity of the training dataset. Further, despite the fact that relatively few subjects and MRI scans were included in our study, the segmentation accuracy growth saturated quickly with increasing size of the training dataset (Fig. 4). This suggests that the biological variability may be a more important characteristic of the training dataset as compared to its size. Hence, clinical implementation of the technique may require additional training of the network on a more heterogeneous dataset. The enhanced training would benefit from including patients with varying age and body mass, the biological factors contributing to structural and image contrast variabilities. Further, the method's performance in clinical cases may be potentially improved by explicit inclusion of lesions and anatomical structures other than cartilage into the labeled training database as previously described[27].

The current study considered computer-assisted manual segmentation as a ground truth for the dataset labeling purposes. Our results demonstrate that the implemented procedure was sufficiently reproducible, with an inter-observer DSC reaching 0.88±0.04 (O1 and O2). This performance is similar to the inter-observer DSC reported for the manual segmentation of knee cartilage (0.88[36]), which is less challenging to segment than cartilage in wrist due to higher thickness and anatomical complexity of the former. Yet, the agreement between O1 and O3 was somewhat lower (DSC = 0.78±0.06). The discordance may be partially explained by differences in training and inter-institutional differences (Observer 3 was affiliated with a different institution than Observers 1 and 2). At the same time, the observation suggests that merging multi-institutional labeled datasets into a single training dataset could be an appropriate training strategy to avoid the bias caused by the segmentation practices adapted by each individual observer and/or the research site.

Our study demonstrated an heterogenous performance of the segmentation with respect to the slice location (Table 2). The best performance was achieved for medial cross-sections of the wrist, in which cartilage tissue is characterized by the most well-defined and inter-connected geometry. The worst performance was observed in the slices located away from the medial cross-section, in which amount of cartilage is low and it may become less recognizable by the network due to poorly defined morphological features. Considering that the cross-validation did not significantly improve the segmentation quality at the periphery relatively to the medial zone (Table 2), we suggest that increasing the number of subjects included in the training dataset may not be the most efficient strategy to improve performance in those locations. At the same time, it is worth noting that increasing the number of slices chosen from each 3D image was not studied separately for lateral slices and should be a subject for a further investigation. The major improvement may come from exploiting 3D peculiarities of cartilage, which are different compared to skin or vessels. Exploting such spatial correlations would require expanding the network architecture into 3D[37]. However, the challenges of such modification are related to a very high computation and memory cost of the 3D-patch-based approach[38] and implementation of 3D image labeling.

**CONCLUSIONS**

The proposed here patch-based CNN-based segmentation of wrist cartilage from MR images provided a time-efficient alternative to the manual segmentation. The proposed architecture outperformed a state-of-the art image-based U-Net architecture and its patch-based variant when trained on a limited amount of subjects' data. Our results highlight the importance of including sufficient number of patients in the training dataset. The accuracy of the proposed approach might be further increased with a more heterogeneous multi-institutional training sample and using a 3D CNN architecture.

**FUNDING SOURCES**

This work was supported by the Russian Science Foundation (Grant No. 18-79-10167). This project has received funding from the European Union's Horizon 2020 research and innovation programme under grant agreement No. 736937. This work was partially supported by NIH (R01EB027087).

**Table 1. Statistics for manual and PB-CNN-assisted segmentation procedures. Quantitative and qualitative validation of methods.**

| # of scan | SNR | Intraobserver DSC | Interobserver DSC (O1 – O2) | Interobserver DSC (O1 – O3) | Intermethod DSC (O1 – CNN) | Visual evaluation score for CNN |
|---|---|---|---|---|---|---|
| 1 | 14.0 | 0.91 | 0.92 | 0.83 | 0.83 | 8 |
| 2 | 19.0 | 0.94 | 0.84 | 0.75 | 0.81 | 7 |
| 3 | 12.4 | 0.89 | 0.90 | 0.83 | 0.83 | 7 |
| 4 | 15.7 | 0.90 | 0.86 | 0.73 | 0.77 | 4 |
| 5 | 14.5 | 0.85 | 0.83 | 0.75 | 0.84 | 7 |
| 6 | 19.6 | 0.92 | 0.81 | 0.75 | 0.82 | 7 |
| 7 | 13.8 | 0.87 | 0.92 | 0.81 | 0.88 | 9 |
| 8 | 20.8 | 0.93 | 0.88 | 0.82 | 0.84 | 7 |
| 9 | 13.7 | 0.97 | 0.91 | 0.81 | 0.86 | 8 |
| 10 | 18.9 | 0.98 | 0.87 | 0.77 | 0.83 | 9 |
| 11 | 12.0 | 0.82 | 0.89 | 0.66 | 0.82 | 7 |
| 12 | 14.2 | 0.96 | 0.89 | 0.74 | 0.84 | 9 |
| 13 | 20.6 | 0.89 | 0.89 | 0.83 | 0.81 | 5 |
| 14 | 10.2 | 0.91 | 0.86 | 0.64 | 0.74 | 5 |
| 15 | 15.1 | 0.87 | 0.94 | 0.85 | 0.81 | 6 |
| *16* | *13.4* | *0.90* | *0.91* | *0.79* | *0.79* | *6* |
| *17* | *21.0* | *0.83* | *0.89* | *0.82* | *0.75* | *5* |
| *18* | *18.6* | *0.93* | *0.92* | *0.83* | *0.80* | *4* |
| *19* | *12.5* | *0.86* | *0.81* | *0.83* | *0.77* | *5* |
| *20* | *15.4* | *0.96* | *0.82* | *0.72* | *0.69* | *3* |
| HV group Mean± SD | **15.57±3.37** | **0.91±0.04** | **0.88±0.03** | **0.77±0.06** | **0.82±0.03** | **7.00±1.51** |
| P group Mean± SD | *16.18±3.57* | *0.90±0.05* | *0.87±0.05* | *0.80±0.04* | *0.76±0.05* | *4.60±1.14* |
| All Groups Mean± SD | 15.7±3.34 | 0.90±0.04 | 0.88±0.04 | 0.78±0.06 | 0.81±0.05 | 6.40±1.76 |

HV - healthy volunteers (MRI scans #1-#15), *P - patients (MRI scans #16-#20),* DSC – Dice coefficient, SD – standard deviation, CNN – convolutional neural network, SNR – signal-to-noise ratio, O1 – observer 1, O2 – observer 2, O3– observer 3, O4 – observer 4.

**Table 2. Results of layer-to-layer analysis of CNN segmentation performance - DSC averaged over zones for "hold-out" and cross-validation (in brackets) studies**

| Subjects group | No cartilage Zone#1 | 1% – 33% Zone#2 | 34% – 66% Zone#3 | 67% – 100% Zone#4 | 3D DSC |
|---|---|---|---|---|---|
| Averaged over HV group ±SD | 0,24 (0.28) ±0,20 (0.20) | 0,66 (0.66) ±0,04 (0.04) | 0,67 (0.68) ±0,05 (0.05) | 0,76 (0.77) ±0,03 (0.03) | 0,73 (0.73) ±0,03 (0.02) |
| *Averaged over P group* ±SD | *0,18 (0.22) ±0,24 (0.21)* | *0,54 (0.56) ±0,08 (0.11)* | *0,58 (0.61) ±0,03 (0.03* | *0,69 (0.72) ±0,05 (0.05)* | *0,65 (0.67) ±0,05 (0.05)* |
| Averaged over all scans ±SD | 0.21 (0.25) ±0.21 (0.20) | 0.60 (0.61) ±0.09 (0.09) | 0.63 (0.65) ±0.06 (0.05) | 0.73 (0.74) ±0.05 (0.05) | 0.69 (0.70) ±0.06 (0.05) |

HV - healthy volunteers (MRI scans #11-#15), *P - patients (MRI scans #16-#20),* DSC – Dice coefficient, CNN – convolutional neural network, SD – Standard deviation.

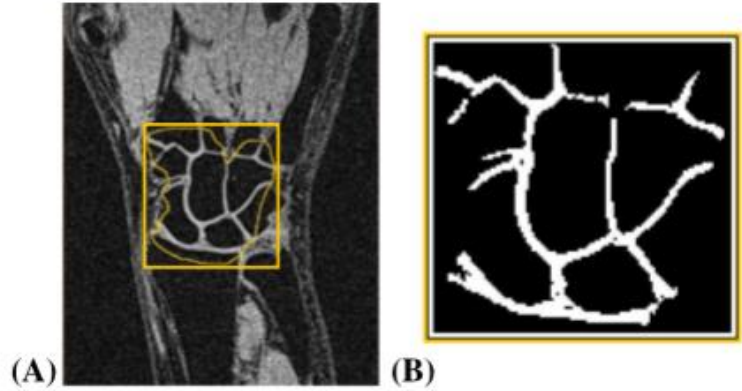

**FIGURE 1** Illustration of the manual segmentation results. (A) Preliminary delineation of the wrist joint area. (B) Final binary mask obtained after thresholding and manual correction

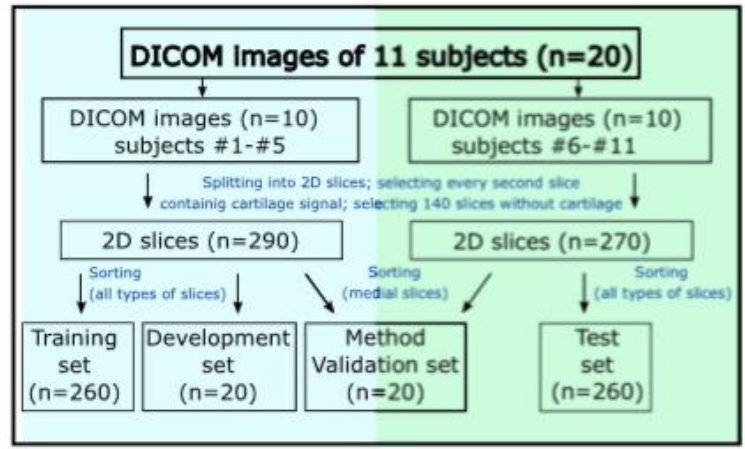

**FIGURE 2** Schematic representation of data splitting for the different stages of the CNN development for the "hold-out" training/testing approach. DICOM, Digital Imaging and Communications in Medicine

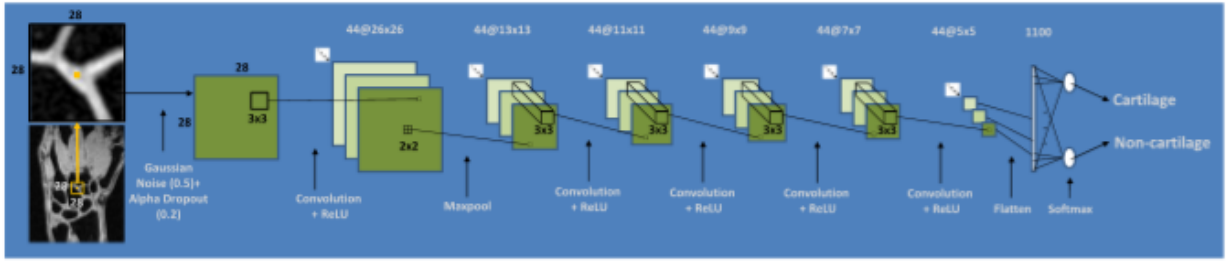

**FIGURE 3** Configuration of PB-CNN optimized for wrist cartilage segmentation

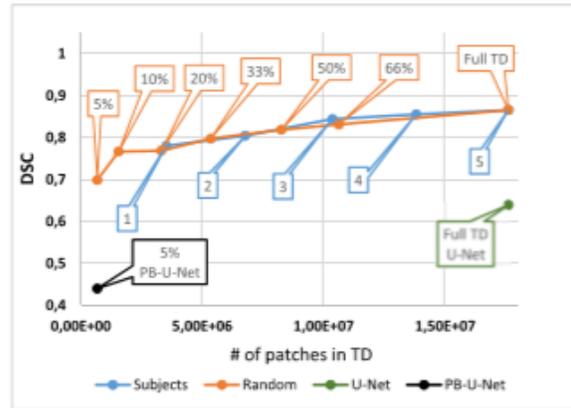

**FIGURE 4** Dependence of DSC value on the training data amount and sample selection ("hold-out" training/testing approach). Blue dots correspond to consecutive inclusion of the data of each subject (from #1 to #5) to the training dataset (TD). Orange dots correspond to a random selection of slices for training in the indicated proportions from the whole TD (full TD). PB-U-Net, patch-based U-Net

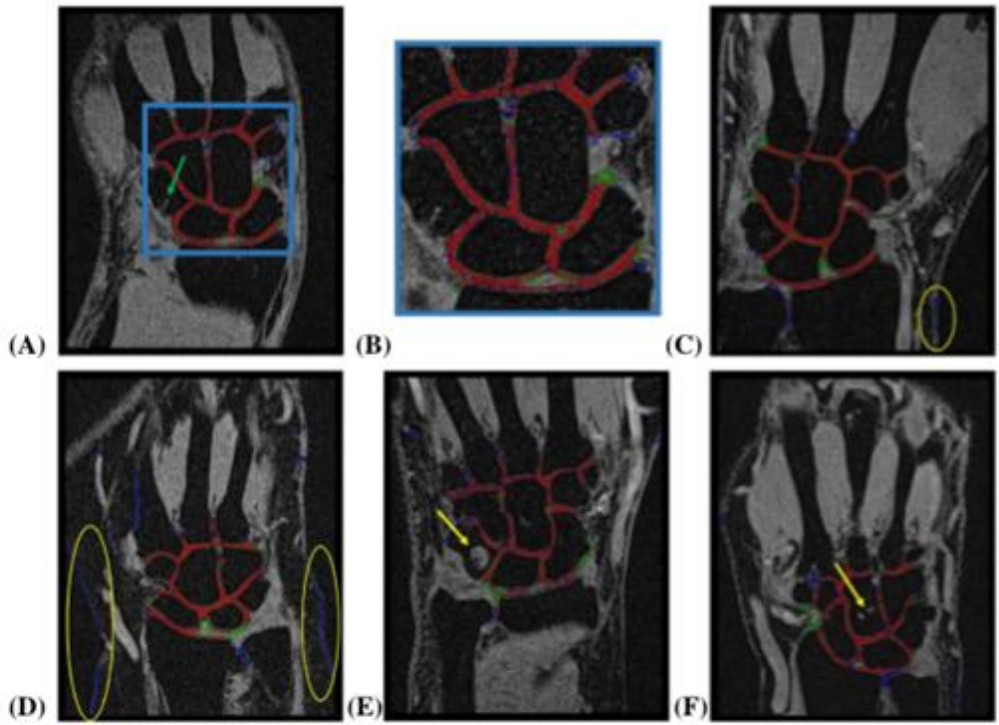

**FIGURE 5** Illustrations of performance of the proposed PB-CNN (red: correctly segmented pixels [true positives]; green: pixels incorrectly assigned to the background [false negatives]; and blue: pixels incorrectly assigned to the cartilage [false positives]). (A) Representative segmentation example (healthy subject, medial slice, DSC = 0.86, visual evaluation score = 8); zoomed-in cartilage area is shown in (B). The green arrow points to a vessel that had contrast and geometry similar to cartilage but was not assigned to this type of tissue by our PB-CNN. (C) Additional segmentation example (healthy subject, medial slice, DSC = 0.81, visual evaluation score = 6). (D) Example of segmentation of patient data with diminished performance (medial slice, DSC = 0.69, visual evaluation score = 3). The yellow circles show the skin tissue considered by CNN as cartilage. (E, F) Additional illustrations of CNN performance on the images of patients. The yellow arrows point to the high signal intensity lesions, which were correctly excluded by the proposed PB-CNN from the segmented mask

# APPENDIX A

FIGURE A1  Summary of the PB-CNN architecture optimized for wrist cartilage segmentation

| Layer (type) | Output Shape | Param # |
|---|---|---|
| gaussian_noise_2 (GaussianNo | (None, 28, 28, 1) | 0 |
| alpha_dropout_2 (AlphaDropou | (None, 28, 28, 1) | 0 |
| conv2d_6 (Conv2D) | (None, 26, 26, 44) | 440 |
| max_pooling2d_2 (MaxPooling2 | (None, 13, 13, 44) | 0 |
| conv2d_7 (Conv2D) | (None, 11, 11, 44) | 17468 |
| conv2d_8 (Conv2D) | (None, 9, 9, 44) | 17468 |
| conv2d_9 (Conv2D) | (None, 7, 7, 44) | 17468 |
| conv2d_10 (Conv2D) | (None, 5, 5, 44) | 17468 |
| flatten_2 (Flatten) | (None, 1100) | 0 |
| dense_2 (Dense) | (None, 2) | 2202 |

Total params: 72,514
Trainable params: 72,514
Non-trainable params: 0

| Layer (type) | Output Shape | Param # | Connected to |
|---|---|---|---|
| input_5 (InputLayer) | (None, 272, 256, 1) | 0 | |
| lambda_5 (Lambda) | (None, 272, 256, 1) | 0 | input_5[0][0] |
| conv2d_73 (Conv2D) | (None, 272, 256, 64) | 128 | lambda_5[0][0] |
| conv2d_74 (Conv2D) | (None, 272, 256, 64) | 4160 | conv2d_73[0][0] |
| max_pooling2d_17 (MaxPooling2D) | (None, 136, 128, 64) | 0 | conv2d_74[0][0] |
| conv2d_75 (Conv2D) | (None, 136, 128, 128) | 8320 | max_pooling2d_17[0][0] |
| conv2d_76 (Conv2D) | (None, 136, 128, 128) | 16512 | conv2d_75[0][0] |
| max_pooling2d_18 (MaxPooling2D) | (None, 68, 64, 128) | 0 | conv2d_76[0][0] |
| conv2d_77 (Conv2D) | (None, 68, 64, 256) | 33024 | max_pooling2d_18[0][0] |
| conv2d_78 (Conv2D) | (None, 68, 64, 256) | 65792 | conv2d_77[0][0] |
| max_pooling2d_19 (MaxPooling2D) | (None, 34, 32, 256) | 0 | conv2d_78[0][0] |
| conv2d_79 (Conv2D) | (None, 34, 32, 512) | 131584 | max_pooling2d_19[0][0] |
| conv2d_80 (Conv2D) | (None, 34, 32, 512) | 262656 | conv2d_79[0][0] |
| max_pooling2d_20 (MaxPooling2D) | (None, 17, 16, 512) | 0 | conv2d_80[0][0] |
| conv2d_81 (Conv2D) | (None, 17, 16, 1014) | 520182 | max_pooling2d_20[0][0] |
| conv2d_transpose_17 (Conv2DTran | (None, 34, 32, 512) | 2077184 | conv2d_81[0][0] |
| concatenate_17 (Concatenate) | (None, 34, 32, 1024) | 0 | conv2d_transpose_17[0][0] conv2d_80[0][0] |

| Layer (type) | Output Shape | Param # | Connected to |
|---|---|---|---|
| conv2d_82 (Conv2D) | (None, 34, 32, 512) | 524800 | concatenate_17[0][0] |
| conv2d_83 (Conv2D) | (None, 34, 32, 512) | 262656 | conv2d_82[0][0] |
| conv2d_transpose_18 (Conv2DTran | (None, 68, 64, 256) | 524544 | conv2d_83[0][0] |
| concatenate_18 (Concatenate) | (None, 68, 64, 512) | 0 | conv2d_transpose_18[0][0] conv2d_78[0][0] |
| conv2d_84 (Conv2D) | (None, 68, 64, 256) | 131328 | concatenate_18[0][0] |
| conv2d_85 (Conv2D) | (None, 68, 64, 256) | 65792 | conv2d_84[0][0] |
| conv2d_transpose_19 (Conv2DTran | (None, 136, 128, 128) | 131200 | conv2d_85[0][0] |
| concatenate_19 (Concatenate) | (None, 136, 128, 256) | 0 | conv2d_transpose_19[0][0] conv2d_76[0][0] |
| conv2d_86 (Conv2D) | (None, 136, 128, 128) | 32896 | concatenate_19[0][0] |
| conv2d_87 (Conv2D) | (None, 136, 128, 128) | 16512 | conv2d_86[0][0] |
| conv2d_transpose_20 (Conv2DTran | (None, 272, 256, 64) | 32832 | conv2d_87[0][0] |
| concatenate_20 (Concatenate) | (None, 272, 256, 128) | 0 | conv2d_transpose_20[0][0] conv2d_74[0][0] |
| conv2d_88 (Conv2D) | (None, 272, 256, 64) | 8256 | concatenate_20[0][0] |
| conv2d_89 (Conv2D) | (None, 272, 256, 64) | 4160 | conv2d_88[0][0] |
| conv2d_90 (Conv2D) | (None, 272, 256, 1) | 65 | conv2d_89[0][0] |

```
Total params: 4,854,583
Trainable params: 4,854,583
Non-trainable params: 0
```

**FIGURE A2** Summary of the basic image-based-U-Net CNN architecture consedered as the state-of-the-art architecture for medical image segmentations

```
Layer (type)                    Output Shape          Param #     Connected to
==================================================================================================
input_1 (InputLayer)            (None, 28, 28, 1)     0
__________________________________________________________________________________________________
lambda_1 (Lambda)               (None, 28, 28, 1)     0           input_1[0][0]
__________________________________________________________________________________________________
conv2d_1 (Conv2D)               (None, 28, 28, 64)    128         lambda_1[0][0]
__________________________________________________________________________________________________
conv2d_2 (Conv2D)               (None, 28, 28, 64)    4160        conv2d_1[0][0]
__________________________________________________________________________________________________
max_pooling2d_1 (MaxPooling2D)  (None, 14, 14, 64)    0           conv2d_2[0][0]
__________________________________________________________________________________________________
conv2d_3 (Conv2D)               (None, 14, 14, 128)   8320        max_pooling2d_1[0][0]
__________________________________________________________________________________________________
conv2d_4 (Conv2D)               (None, 14, 14, 128)   16512       conv2d_3[0][0]
__________________________________________________________________________________________________
max_pooling2d_2 (MaxPooling2D)  (None, 7, 7, 128)     0           conv2d_4[0][0]
__________________________________________________________________________________________________
conv2d_5 (Conv2D)               (None, 7, 7, 256)     33024       max_pooling2d_2[0][0]
__________________________________________________________________________________________________
conv2d_transpose_1 (Conv2DTrans (None, 14, 14, 128)   131200      conv2d_5[0][0]
__________________________________________________________________________________________________
concatenate_1 (Concatenate)     (None, 14, 14, 256)   0           conv2d_transpose_1[0][0]
                                                                  conv2d_4[0][0]
__________________________________________________________________________________________________
conv2d_6 (Conv2D)               (None, 14, 14, 128)   32896       concatenate_1[0][0]
__________________________________________________________________________________________________
conv2d_7 (Conv2D)               (None, 14, 14, 128)   16512       conv2d_6[0][0]
__________________________________________________________________________________________________
conv2d_transpose_2 (Conv2DTrans (None, 28, 28, 64)    32832       conv2d_7[0][0]
__________________________________________________________________________________________________
concatenate_2 (Concatenate)     (None, 28, 28, 128)   0           conv2d_transpose_2[0][0]
                                                                  conv2d_2[0][0]
__________________________________________________________________________________________________
conv2d_8 (Conv2D)               (None, 28, 28, 64)    8256        concatenate_2[0][0]
__________________________________________________________________________________________________
conv2d_9 (Conv2D)               (None, 28, 28, 64)    4160        conv2d_8[0][0]
__________________________________________________________________________________________________
conv2d_10 (Conv2D)              (None, 28, 28, 1)     65          conv2d_9[0][0]
==================================================================================================
Total params: 288,065
Trainable params: 288,065
Non-trainable params: 0
```

**FIGURE A3** Summary of the basic patch-based-U-Net CNN architecture consedered as the state-of-the-art patch-based architecture for medical image segmentations